# Pilot Contamination Mitigation for Wideband Massive MMO: Number of Cells Vs Multipath

Tadilo Endeshaw Bogale[+], Long Bao Le[+], Xianbin Wang[++] and Luc Vandendorpe[+++]
Institute National de la Recherche Scientifique (INRS), Canada[+]
University of Western Ontario, Canada[++]
University Catholique de Louvain, Belgium[+++]
Email: {tadilo.bogale, long.le}@emt.inrs.ca, xianbin.wang@uwo.ca and luc.vandendorpe@uclouvain.be

*Abstract*—This paper proposes novel joint channel estimation and beamforming approach for multicell wideband massive multiple input multiple output (MIMO) systems. Using our channel estimation and beamforming approach, we determine the number of cells $N_c$ that can utilize the same time and frequency resource while mitigating the effect of pilot contamination. The proposed approach exploits the multipath characteristics of wideband channels. Specifically, when the channel has $L$ multipath taps, it is shown that $N_c \leq L$ cells can reliably estimate the channels of their user equipments (UEs) and perform beamforming while mitigating the effect of pilot contamination. For example, in a long term evolution (LTE) channel environment having delay spread $T_d = 4.69\mu$ second and channel bandwidth $B = 2.5$**MHz**, we have found that $L = 18$ cells can use this band. In practice, $T_d$ is constant for a particular environment and carrier frequency, and hence $L$ increases as the bandwidth increases. The proposed channel estimation and beamforming design is linear, simple to implement and significantly outperforms the existing designs, and is validated by extensive simulations.

*Index Terms*— Beamforming, Channel estimation, Massive MIMO, Pilot contamination, Rayleigh Quotient.

## I. INTRODUCTION

Massive multiple input and multiple output (MIMO) is one of the promising techniques to satisfy the gigabit connectivity requirements of the future 5G network [1]. In a multiuser and multicell setup, massive MIMO can be deployed both at the base stations (BSs) and user equipments (UEs). However, since the energy consumption and cost of transceiver devices increase as the number of antennas increases, deploying massive antennas at the UEs is usually infeasible from practical aspect particularly at microwave frequency bands. This motivates the current paper to consider a massive MIMO system where each BS (UE) has massive antennas (single antenna) as in [2], [3].

The potentials of a massive MIMO system can be exploited by performing beamforming which depends on the availability of accurate channel state information (CSI). However, in practice, the channel between each BS and UE is estimated from orthogonal pilot sequences which are limited by the coherence time of the channel [3]–[5]. For a massive MIMO system, one can possibly estimate and exploit the maximum number of channel coefficients using time division duplex (TDD) approach by sending pilots from UEs [5], [6]. By doing so, each BS will estimate, and utilize the channel coefficients both for the downlink and uplink channel data transmissions. In a multicell setup where there is no coordination between cells, each BS determines its number of UEs from the coherence time of the channel only and the UEs in each cell utilizes the same pilot sequences. The reuse of pilot sequences of several co-channel cells will create a so called *pilot contamination* which is shown to be a fundamental performance bottleneck for multicell massive MIMO systems [1]–[3], [5]–[9].

A number of approaches are proposed to address pilot contamination. Eigenvalue decomposition (EVD) based channel estimation approach is proposed in [7] where it is demonstrated that this method achieves better estimation accuracy than that of the conventional least square (LS) and minimum mean square error (MMSE) estimators. A coordinated channel estimation approach is proposed in [2] where it is shown that pilot contamination can be vanished when UEs have special channel covariance matrices. In [8], pilot contamination elimination approach is proposed by allowing pilot transmissions both in the uplink and downlink channels. Successive pilot transmission approach has been proposed in [9] to eliminate pilot contamination. This paper utilizes consecutive pilot transmission phases in which each BS stays idle at one phase and repeatedly transmits pilot sequences in other phases.

In [4], [10], time shifted channel estimation approach for orthogonal frequency division multiplexing (OFDM) based frequency selective channels is proposed. In such approach, first the cells are grouped in some predefined way. Then, only one group of cells send pilot signal sequentially while all the other groups of cells transmit their downlink data signal. Finally, all cells transmit their uplink data on the remaining time slot. In this approach, however, since the number of cells in each group could be more than one, large coherence time is required to cancel pilot contamination in each group of cells when every cell serves the same number of UEs like in the conventional LS and MMSE approaches (for example like in [3], [4]). From these discussions, one can notice that the problem of pilot contamination has been considered by several papers. However, some of the aforementioned papers address pilot contamination by exploiting particular channel covariance information which may not be hold in all scenarios, and the rest of them utilize extra pilots compared to [3], [4].

The current paper examines a pilot contamination problem for multicell massive MIMO systems with frequency selective channels. For a given pilot duration, we assume that the number of UEs served by each BS is the same as that of [3], [4]. Furthermore, the channel covariance matrix of each

UE is assumed to have "arbitrary structure". Under these settings and assumptions, we propose a novel joint channel estimation and beamforming approach. The proposed approach exploits the inherent multipath behavior of frequency selective channels. Specifically, when the channel has $L$ multipath taps, it is shown that $N_c \leq L$ cells can reliably estimate the channels of their UEs and perform beamforming while efficiently mitigating the effect of pilot contamination.

In other words, when $N_c \leq L$, it is shown that the signal to interference plus raise ratio (SINR) of each UE's sub-carrier increases as the number of antennas at each BS $N$ increases (i.e., the SINR grows indefinitely as $N \to \infty$). In an exemplify long term evolution (LTE) channel environment having delay spread $T_d = 4.69\mu$ seconds and channel bandwidth $B = 2.5$MHz, $L = 18$ cells can use this frequency band while efficiently mitigating the effect of pilot contamination. In practice, $T_d$ is constant for a particular environment and carrier frequency, and hence $L$ increases as the bandwidth increases [11]. The proposed design is linear, simple to implement and significantly outperforms the existing designs, and is validated by extensive simulations.

This paper is organized as follows. Section II discusses the system and channel models. In Section III, a summary of pilot contamination and the objective of the paper is explained. In Sections IV - VI, the proposed algorithm and its performance analysis is presented briefly. In Section VII, simulation results are presented. Finally, Section VIII draws conclusions.

*Notations:* In this paper, upper/lower-case boldface letters denote matrices/column vectors of appropriate size. Identity matrix, singular value decomposition, expectation and absolute value are denoted by $\mathbf{I}$, SVD(.), E(.) and |.|, respectively. Greater than or approximately equal to is denoted by $\gtrapprox$.

## II. SYSTEM AND CHANNEL MODEL

We consider a multiuser and multicell system with $N_c$ cells (i.e., BSs) where each UE and BS are equipped with 1 and $N$ antennas, respectively. We consider transmission scheme with symbol period $T_s$ and channel environment with maximum delay spread $T_d$[1]. The total transmission bandwidth becomes $B = B_0(1+\alpha)$, where $B_0 = \frac{1}{2T_s}$ is the Nyquist bandwidth and $\alpha$ is the excess bandwidth [13], [14][2]. For these settings, the number of multipath channel taps $L$ is approximated as $L = \frac{T_d}{T_s}$ [11] (see page 143) and the channel coefficients between the $k$th UE in $j$th cell to the $i$th BS $n$th antenna can be denoted as [12], [14]

$$\bar{\mathbf{h}}_{kjin} = [\bar{h}_{kjin1}, \bar{h}_{kjin2}, \cdots, \bar{h}_{kjinL}]^T \quad (1)$$

where $\bar{h}_{kjinl} = \sqrt{g_{kji}}\tilde{h}_{kjinl}$ with $\tilde{h}_{kjinl}, \forall k, j, i, n, l$ ($g_{kji}$) are fast fading (distance dependent slow fading) coefficients.

In most systems having multipath channels, OFDM based transmission is adopted. For such a transmission, the channel coefficient of each sub-carrier has practical importance. The channel coefficient between the $k$th UE in $j$th cell to the $i$th BS $n$th antenna in sub-carrier $s$ can be computed from (1) as

$$h_{kjins} = \mathbf{f}_s^H \bar{\mathbf{h}}_{kjin} \quad (2)$$

where $\mathbf{f}_s^H = [1, e^{-\sqrt{-1}\frac{2\pi}{M}s}, e^{-\sqrt{-1}\frac{2\pi}{M}2s}, \cdots, e^{-\sqrt{-1}\frac{2\pi}{M}(L-1)s}]$ with $M$ as the fast Fourier transform (FFT) size of the OFDM.

## III. PILOT CONTAMINATION AND OBJECTIVE

The number of terminals served by a BS is proportional to the pilot duration $T_p$. For the given $T_p$, the channel coefficients of each UE can be estimated with or without OFDM pilot transmission. By denoting each OFDM duration as $T_o$ and useful symbol duration as $T_u$, each BS can serve the following maximum number of UEs (see (4) of [3] and Section II of [4]):

$$\frac{T_p}{T_d}\frac{T_u}{T_o} \leq \frac{T_p}{T_d} = \frac{N_p}{L} \triangleq \tilde{K} \quad (3)$$

where $N_p = \frac{T_p}{T_s}$ is the number of samples acquired in $T_p$. If each UE applies non OFDM pilot transmission, it is possible to serve $\tilde{K}$ UEs [3].

If a BS serves $\tilde{K}$ UEs (i.e., use $T_p$ to estimate the channel coefficients of these UEs), we can understand from the work of [3] that only one cell can serve its UEs without experiencing any pilot contamination. In particular, when $\tilde{h}_{kjinl}, \forall k, j, i, n, l$ are modeled as independent and identically distributed (i.i.d) zero mean circularly symmetric complex Gaussian (ZMCSCG) random variables each with unit variance (i.e., $\tilde{h}_{kjinl} \sim \mathcal{CN}(0,1), \forall k, j, i, n, l$). And if there are multiple cells where the BS of each cell has sufficiently large number of antennas (i.e., $N \to \infty$), the effect of noise vanishes and the SINR of the $k$th UE in cell $i$ and sub-carrier $s$ is given as

$$SINR_{kis} = \frac{E\{h_{kiins}^H h_{kiins}\}}{\sum_{j \neq i}^{N_c} E\{h_{kjins}^H h_{kjins}\}} = \frac{g_{kii}^2}{\sum_{j \neq i}^{N_c} g_{kij}^2}. \quad (4)$$

As we can see from this equation, $SINR_{kis}$ is bounded even if $N \to \infty$. However, one can understand from MIMO communication that increasing $N$ should help to improve $SINR_{kis}$ if pilot contamination is mitigated or cancelled. This motivates us to consider the following objective:
For the given $B$, $T_p$, $T_o$, $T_s$, $T_d$ and $T_u$, the channel vector between any two UEs are uncorrelated and each BS serves $\tilde{K}$ UEs (i.e., the same setting as in [3], [4]), how many cells can utilize the same time-frequency resource while ensuring that $SINR_{kis}$ increases as $N$ increases (i.e., achieving unbounded $SINR_{kis}$ as $N \to \infty$ which is the same as mitigating or possibly canceling the effect of pilot contamination)?

## IV. PROPOSED CHANNEL ESTIMATION AND BEAMFORMING

This section discusses the proposed channel estimation and beamforming approach. In this regard, we assume that the pilots are transmitted without OFDM whereas, the data symbols are transmitted using OFDM scheme. The current paper considers that both pilot and data transmissions take place in the uplink channel. We also assume that $N_c$ (which will be determined in the sequel) cells utilize the same time-frequency resource while mitigating the effect of pilot contamination. Here, we consider that the $i$th cell is the target cell.

---
[1]The transmission environment has a maximum delay spread of $T_d$ which has typical values $T_d = \{4.7, 5.2\}\mu s$ for urban cells and $T_d = 16.7\mu s$ for very large multi-cell MIMO systems in the existing LTE network [3], [12].

[2]The exact value of $\alpha$ varies from one application to another (for example, $0.2 \leq \alpha \leq 0.35$ is used in [13]).

## A. Channel Estimation

This subsection provides the proposed channel estimation. For the considered settings, the received signal at the $n$th antenna of the $i$th BS can be expressed as

$$\mathbf{r}_{in} = \sum_{k=1}^{\tilde{K}}(\mathbf{X}_{ki}\bar{\mathbf{h}}_{kiin} + \sum_{j=1,j\neq i}^{N_c}\mathbf{X}_{kj}\bar{\mathbf{h}}_{kjin}) + \mathbf{w}_{in}$$
$$= \mathbf{X}_i\bar{\mathbf{h}}_{iin} + \tilde{\mathbf{X}}_i\tilde{\mathbf{h}}_{iin} + \mathbf{w}_{in} \quad (5)$$

where $\bar{\mathbf{h}}_{jin} = [\bar{\mathbf{h}}_{1jin}^T, \bar{\mathbf{h}}_{2jin}^T, \cdots, \bar{\mathbf{h}}_{\tilde{K}jin}^T]^T$, $\tilde{\mathbf{h}}_{iin} = [\bar{\mathbf{h}}_{1in}^T, \bar{\mathbf{h}}_{2in}^T, \cdots, \bar{\mathbf{h}}_{(i-1)in}^T, \bar{\mathbf{h}}_{(i+1)in}^T, \cdots, \bar{\mathbf{h}}_{N_cin}^T]^T$, $\tilde{\mathbf{X}}_i = [\mathbf{X}_1, \mathbf{X}_2, \cdots, \mathbf{X}_{i-1}, \mathbf{X}_{i+1}, \cdots, \mathbf{X}_{N_c}] \in \mathcal{C}^{N_p \times (N_p(N_c-1))}$, $\mathbf{X}_i = [\mathbf{X}_{1i}, \mathbf{X}_{2i}, \cdots, \mathbf{X}_{\tilde{K}i}]$ and $\mathbf{w}_{in} \in \mathcal{C}^{N_p \times 1}$ is the additive noise vector during pilot transmission phase where its entries are assumed to be i.i.d ZMCSCG random variables each with variance $\sigma^2$, and $\mathbf{X}_{kj} \in \mathcal{C}^{N_p \times L}$ is a Toeplitz matrix, i.e.,

$$\mathbf{X}_{kj} = \begin{bmatrix} x_{kj_1} & 0 & \cdots & 0 & 0 \\ x_{kj_2} & x_{kj_1} & \cdots & \vdots & \vdots \\ x_{kj_3} & x_{kj_2} & \cdots & 0 & 0 \\ \vdots & \vdots & \cdots & \vdots & \vdots \\ x_{kj_{(N_p-1)}} & x_{kj_{(N_p-2)}} & \cdots & x_{kj_{(N_p-L+1)}} & x_{kj_{(N_p-L)}} \\ x_{kj_{N_p}} & x_{kj_{N_p-1}} & \cdots & x_{kj_{(N_p-L+2)}} & x_{kj_{(N_p-L+1)}} \end{bmatrix} \quad (6)$$

with $\mathbf{x}_{kj} = [x_{kj_1}, x_{kj_2} \cdots, x_{kj_{N_p}}]^T$ as the pilot symbols transmitted from the $k$th UE in cell $j$ which will be designed in Section V. As can be seen from (5), the $i$th BS experiences pilot contamination due to the term $\tilde{\mathbf{X}}_i\tilde{\mathbf{h}}_{iin}$ [3]. For an arbitrary pilot, one can also notice from this equation that $\tilde{\mathbf{X}}_i\tilde{\mathbf{h}}_{iin} = 0$ is ensured when $N_c = 1$ (i.e., like in [3], [4]). For such setting, one can apply MMSE or LS method to estimate $\bar{\mathbf{h}}_{iin}$ from $\mathbf{r}_{in}$.

In the following, we provide the proposed channel estimation approach. In this regard, we introduce a vector $\mathbf{v}_{kis} \in \mathcal{C}^{N_p \times 1}$ to estimate the $s$th sub-carrier channel of the $k$th UE in cell $i$ which will be designed in Section V. Using this linear combination vector, we express the estimate of $h_{kiins}$ as

$$\hat{h}_{kiins} = \mathbf{r}_{in}^T\mathbf{v}_{kis} = (\mathbf{X}_i\bar{\mathbf{h}}_{iin} + \sum_{j=1,j\neq i}^{N_c}\mathbf{X}_j\bar{\mathbf{h}}_{jin} + \mathbf{w}_{in})^T\mathbf{v}_{kis}. \quad (7)$$

As will be clear in the next section, this estimated channel helps us to increase the number of cells more than one.

## B. Beamforming

As discussed above, we consider the uplink channel for data transmission using OFDM approach. Since data transmission takes place using OFDM technique, the received signal of each sub-carrier can be obtained independently at each antenna of all BSs. To this end, the $i$th BS $n$th antenna receives the following uplink signal in sub-carrier $s$ ($y_{ins}$)

$$y_{ins} = \sum_{k=1}^{\tilde{K}}\sum_{j=1}^{N_c}h_{kjins}d_{kjs} + \tilde{w}_{ins}$$

where $\tilde{w}_{ins}$ is the noise sample at the $i$th BS $n$th antenna and sub-carrier $s$ during data transmission and $d_{kjs}$ is the data symbol transmitted in sub-carrier $s$ of the $k$th UE in cell $j$. The overall received signal at the $i$th BS $s$th sub-carrier becomes

$$\mathbf{y}_{is} = \sum_{k=1}^{\tilde{K}}\sum_{j=1}^{N_c}\mathbf{h}_{kjis}d_{kjs} + \tilde{\mathbf{w}}_{is}$$
$$= \mathbf{h}_{kiis}d_{kis} + \sum_{m=1}^{\tilde{K}}\sum_{j=1,(m,j)\neq(k,i)}^{N_c}\mathbf{h}_{mjis}d_{mjs} + \tilde{\mathbf{w}}_{is}$$

where $\mathbf{y}_{is} = [y_{i1s}, y_{i2s}, \cdots, y_{iNs}]^T$ and $\tilde{\mathbf{w}}_{is} = [\tilde{w}_{i1s}, \tilde{w}_{i2s}, \cdots, \tilde{w}_{iNs}]^T$. It is assumed that each entry of $\tilde{\mathbf{w}}_{is}$ is assumed to be i.i.d ZMCSCG random variable with variance $\tilde{\sigma}^2$. Now, let us assume that we are interested to estimate $d_{kis}$ using a beamforming vector $\mathbf{a}_{kiis} \in \mathcal{C}^{N \times 1}$ as

$$\hat{d}_{kis} = \mathbf{a}_{kiis}^H\mathbf{y}_{is} \quad (8)$$
$$= \mathbf{a}_{kiis}^H(\mathbf{h}_{kiis}d_{kis} + \sum_{m=1}^{\tilde{K}}\sum_{j=1,(m,j)\neq(k,i)}^{N_c}\mathbf{h}_{mjis}d_{mjs} + \tilde{\mathbf{w}}_{is}).$$

Using this beamformer, $\hat{d}_{kis}$ will have the following SINR

$$\bar{\gamma}_{kis} = \frac{\mathrm{E}|\mathbf{h}_{kiis}^H\mathbf{a}_{kiis}|^2}{\sum_{(m,j)\neq(k,i)}\mathrm{E}|\mathbf{h}_{mjis}^H\mathbf{a}_{kiis}|^2 + \mathrm{E}|\tilde{\mathbf{w}}_{is}^H\mathbf{a}_{kiis}|^2}. \quad (9)$$

As can be seen from this equation, the achievable $\bar{\gamma}_{kis}$ depends on $\mathbf{a}_{kiis}$. Furthermore, one can utilize different beamforming methods to design $\mathbf{a}_{kiis}$. In a massive MIMO system, simple beamforming methods such as maximum ratio combining (MRC) and zero forcing (ZF) are close to optimal which motivates us to choose $\mathbf{a}_{kiis}$ to be the MRC beamformer [3], [5], [6]. That is $\mathbf{a}_{kiis} = \hat{\mathbf{h}}_{kiis} = [\hat{h}_{kii1s}, \hat{h}_{kii2s} \cdots, \hat{h}_{kiiNs}]^T$, where $\hat{h}_{kiins}$ is the estimated channel defined in (7). With this beamforming vector, we will have

$$\mathrm{E}|\mathbf{w}_i^H\mathbf{a}_{kiis}|^2 = \mathbf{v}_{kis}^H\left(\sum_{m=1}^{\tilde{K}}\sum_{j=1}^{N_c}\mathbf{X}_{mj}^*\mathbf{C}_{mji}\mathbf{X}_{mj}^T + \sigma^2\mathbf{I}\right)\mathbf{v}_{kis}$$

$$\mathrm{E}|\mathbf{h}_{mjis}^H\mathbf{a}_{kiis}|^2 = \mathbf{v}_{kis}^H\left(\mathbf{X}_{mj}^*\left[N\mathbf{C}_{mji}\mathbf{f}_s^*\mathbf{f}_s^T\mathbf{C}_{mji} + \right.\right. \quad (10)$$
$$\left.\left.\sum_{u=1}^{\tilde{K}}\sum_{v=1,(u,v)\neq(m,j)}^{N_c}L\mathbf{C}_{mjuvis}\right]\mathbf{X}_{mj}^T + \frac{L}{N}\sigma^2\mathbf{C}_{mjis}\right)\mathbf{v}_{kis}$$

with $\mathbf{C}_{mjuvis} = \mathrm{E}\{\bar{\mathbf{H}}_{mji}^*\bar{\mathbf{H}}_{uvi}^T\mathbf{f}_s^*\mathbf{f}_s^T\bar{\mathbf{H}}_{uvi}^*\bar{\mathbf{H}}_{mji}^T\}$ and $\mathbf{C}_{mjis} = \mathrm{E}\{\bar{\mathbf{H}}_{mji}^T\mathbf{f}_s^*\mathbf{f}_s^T\bar{\mathbf{H}}_{mji}^*\}$ (see Appendix A of [15] for the details).

The beamforming vector $\mathbf{a}_{kiis}$ exploits the estimated channel (7) which is a function of $\mathbf{v}_{kis}$, $\mathbf{x}_{mj}$, $\forall m,j$ and $N_c$. Hence, the achievable $\bar{\gamma}_{kis}$ depends on these variables that need to be optimized which is the focus of the next section.

## V. OPTIMIZATION OF $N_c$, $\mathbf{x}_{mj}$ AND $\mathbf{v}_{kis}$

This subsection discusses the optimization of $\mathbf{v}_{kis}$, $\mathbf{x}_{mj}$ and $N_c$. In this regard, we examine two problems: In the first problem, we determine the number of cells $N_c$, and feasible $\mathbf{v}_{kis}$ and $\mathbf{x}_{mj}$ such that the multi-cell system is able to mitigate pilot contamination. In the second problem, we re-optimize $\mathbf{v}_{kis}$ to further maximize $\bar{\gamma}_{kis}$ for fixed $N_c$ and $\mathbf{x}_{mj}$, $\forall m,j$.

## A. Determination of $N_c$, and Feasible $\mathbf{v}_{kis}$ and $\mathbf{x}_{mj}$

This subsection determines $N_c$, and feasible $\mathbf{v}_{kis}$ and $\mathbf{x}_{mj}$ such that the multicell system is able to mitigate the effect of pilot contamination. Specifically, we determine these variables while ensuring that $\bar{\gamma}_{kis}$ increases as $N$ increases (i.e., $\bar{\gamma}_{kis}$ grows to infinity as $N \rightarrow \infty$) in the following Theorem.

*Theorem 1:* For arbitrary channel covariance matrices given in (10), $\bar{\gamma}_{kis}$ of (9) increases as $N$ increases (i.e., the effect of pilot contamination is mitigated) when $N_c \leq L$, $\mathbf{x}_{mj}, \forall m, j$ are selected such that $\mathbf{Q}_{is}^T, \forall i, s$ are full row rank matrices and $\mathbf{v}_{kis}$ is designed from the solution of

$$\mathbf{D}_{is}\bar{\mathbf{U}}_{is}\mathbf{v}_{kis} = \mathbf{u}_{kiis} \qquad (11)$$

where $\mathbf{Q}_{is} = [\mathbf{q}_{11is}, \mathbf{q}_{21is}, \cdots, \mathbf{q}_{\tilde{K}1is}, \mathbf{q}_{12is}, \mathbf{q}_{22is}, \cdots, \mathbf{q}_{\tilde{K}2is}, \cdots, \mathbf{q}_{1N_cis}, \mathbf{q}_{2N_cis}, \cdots, \mathbf{q}_{\tilde{K}N_cis}]$, $\mathbf{q}_{mjis}^T = \mathbf{f}_s^T \mathbf{C}_{mji}\mathbf{X}_{mj}^T$, $\mathbf{U}_{is}^H\mathbf{D}_{is}\bar{\mathbf{U}}_{is} = \text{SVD}(\mathbf{Q}_{is}^T)$, $\mathbf{U}_{is} \in \mathcal{C}^{N_p \times N_p}$ and $\bar{\mathbf{U}}_{is} \in \mathcal{C}^{N_p \times N_p}$ are unitary matrices, $\mathbf{u}_{kiis}^H$ is the row vector of $\mathbf{U}_{is}^H$ corresponding to $\mathbf{q}_{kiis}^T$ and $\mathbf{D}_{is} \in \mathcal{C}^{N_p \times N_p}$ is a diagonal matrix containing the singular values of $\mathbf{Q}_{is}^T$.

*Proof:* See Appendix B of [15]. ∎

For arbitrary covariance matrices $\mathbf{C}_{(.)}$ of (10), one approach of satisfying full rank $\mathbf{Q}_{is}^T$ is by selecting $\mathbf{x}_{mj}, \forall m, j$ so that they will be uncorrelated to each other. Furthermore, in order to maintain full rank $\mathbf{Q}_{is}^T$ for all sub-carriers, $\mathbf{x}_{mj}, \forall m, j$ may need to have a "white noise" like sequence. This motivates us to select $\mathbf{x}_{11}$ from noise like deterministic (or random) sequence, and then construct $\mathbf{x}_{mj}, \forall (m, j) \neq (1, 1)$ by shifting the elements of $\mathbf{x}_{11}$ with $N_p$ sized FFT matrix. With such selection, however, we have noticed that the average rate achieved by one sub-carrier differs from the other sub-carrier and this rate imbalance may not be desirable in practice. For this reason, we choose $\mathbf{x}_{11}$ randomly from a given set of sequences where these sequences are known a priori to all BSs and UEs. Upon doing so, we have observed that each sub-carrier of a UE achieves the same average rate.

One can understand from the result of *Theorem 1* that if $N_c > L$, the proposed channel estimation and beamforming approach will achieve "bounded $\bar{\gamma}_{kis}$" even though $N \rightarrow \infty$ (i.e., if $N_c > L$, we will experience pilot contamination as will be demonstrated in the simulation section).

## B. Re-optimization of $\mathbf{v}_{kis}$

From the above subsection, we are able to determine $N_c$, and provide feasible $\mathbf{v}_{kis}$ and $\mathbf{x}_{mj}$ while ensuring that the effect of pilot contamination is mitigated. For the given $N_c$, it is also possible to optimize $\mathbf{v}_{kis}$ and $\mathbf{x}_{mj}$ to further increase $\bar{\gamma}_{kis}$. However, jointly optimizing $\mathbf{v}_{kis}$ and $\mathbf{x}_{mj}$ to maximize $\bar{\gamma}_{kis}$ is still complicated. For this reason, we re-optimize $\mathbf{v}_{kis}$ only to maximize $\bar{\gamma}_{kis}$ for fixed $N_c$ and $\mathbf{x}_{mj}$ as follows.

$$\max_{\mathbf{v}_{kis}} \bar{\gamma}_{kis}. \qquad (12)$$

This problem is a Rayleigh quotient optimization problem where generalized eigenvalue solution approach can be applied to get the optimal $\mathbf{v}_{kis}$ [13], [16], [17].

The next issue is what is the relation between the solutions obtained from (11) and (12)? In the following, we address these issues: Denote the solution obtained by solving (12) as $\mathbf{v}_{kis}^*$ and its corresponding SINR as $\gamma_{kis}^*$. As this problem is a Rayleigh quotient, $\mathbf{v}_{kis}^*$ is a global optimal solution. Hence, the SINR obtained by any $\mathbf{v}_{kis} \neq \mathbf{v}_{kis}^*$ can not be higher than that of $\gamma_{kis}^*$ which leads to (after some mathematical steps)

$$\gamma_{kis}^* \geq \bar{\gamma}_{kis} \qquad (13)$$

where

$$\bar{\gamma}_{kis} = \frac{N+a}{b} \bigg|_{\mathbf{v}_{kis}= \text{ soln. of } (11)} \qquad (14)$$

where $a = \mathbf{v}_{kis}^H(\sum_{u=1}^{\tilde{K}} \sum_{v=1,(u,v)\neq(k,i)}^{N_c} L\mathbf{X}_{mj}^*\mathbf{C}_{kiuvis}\mathbf{X}_{mj}^T + L\sigma^2\mathbf{C}_{kiis})\mathbf{v}_{kis}$, $b = \mathbf{v}_{kis}^H(\sum_{m=1}^{\tilde{K}} \sum_{j=1}^{N_c} \sigma^2\mathbf{X}_{mj}^*\mathbf{C}_{mji}\mathbf{X}_{mj}^T + \sum_{(m,j)\neq(k,i)} \sum_{(u,v)\neq(m,j)} L\mathbf{X}_{mj}^*\mathbf{C}_{mjuvis}\mathbf{X}_{mj}^T + L\sigma^2\mathbf{C}_{mjis} + \sigma^4\mathbf{I})\mathbf{v}_{kis}$. As can be seen from this expression, $a$ and $b$ are constant terms which are not dependent on $N$. Hence, for fixed $N_c, L$ and $\tilde{K}$, one can expect that the SINR obtained by utilizing $\mathbf{v}_{kis}$ of (11) and (12) will be almost the same for large $N$. However, the solutions of these two problems may have different SINRs when $N$ is not sufficiently large. This fact has been verified in the simulation section.

We would like to stress here that the current paper utilizes pilot transmission without OFDM as shown in (5). However, as clearly seen from this equation, performing pilot transmission without OFDM "alone" does not bring any new result to increase $N_c$. Nevertheless, by performing pilot transmission without OFDM, data transmission with OFDM, carefully selecting $\mathbf{x}_{mj}$, and introducing and optimizing the variable $\mathbf{v}_{kis}$, we are able to increase the number of cells to $N_c = L$.

## VI. Performance Analysis

In a practical wireless system, the rate achieved by $\hat{d}_{kis}$ (i.e., $R_{kis}$) which is related to $\bar{\gamma}_{kis}$ has an interest. Furthermore, $R_{kis}$ will be maximum when each of the BSs has perfect CSI of its own UEs. This motivates us to examine $R_{kis}$ of the proposed approach and that of perfect CSI scenario. In this regard, we consider the following theorem.

*Theorem 2* When $\tilde{h}_{kji}$ are modeled as in (4), $\gamma_{kis}^{pe} \gg 1$, $\gamma_{kis}^{pr} \gg 1$ and $g_{kii} \gtrapprox g_{mji}$ (the case in practice), we have

$$R_g \triangleq R_{kis}^{pe} - R_{kis}^{pr} \leq \log_2(\eta) = c_0, \Rightarrow R_{kis}^{pe} \approx R_{kis}^{pr}\bigg|_{N \rightarrow \infty} \qquad (15)$$

where $R_{kis}^{pe}$ and $R_{kis}^{pr}$ are the rate achieved by perfect CSI scenario and proposed approach, respectively and the approximation is because $\eta$ is independent of $N$ and is given as

$$\eta = \frac{L\tilde{b}}{g_{kii}(\sum_{m=1}^{\tilde{K}} \sum_{j=1,(m,j)\neq(k,i)}^{N_c} Lg_{kji} + \sigma^2)}\bigg|_{\mathbf{v}_{kis}= \text{ soln. of } (11)}$$

with $\tilde{b} = \mathbf{v}_{kis}^H(\sum_{(m,j)\neq(k,i)}[\sum_{(u,v)\neq(m,j)} Lg_{mji}g_{uvi}\mathbf{X}_{mj}^*\mathbf{X}_{mj}^T + L\sigma^2 g_{mji}\mathbf{I}] + \sigma^2(\sum_{(m,j)} g_{mji}\mathbf{X}_{mj}^*\mathbf{X}_{mj}^T + \sigma^2\mathbf{I}))\mathbf{v}_{kis}$.

*Proof:* See Appendix C of [15]. ∎

We would like to mention here that the results of the current paper are provided for the MRC beamforming approach in the uplink channel only. However, the approach of this paper can also be extended for other beamforming techniques and downlink channel. For such cases, $N_c$ and $\mathbf{x}_{mj}$ are the same as that of the current paper. However, the optimal solution of

$\mathbf{v}_{kis}$ may not be necessarily the same for all beamforming types and channels. The detailed extension is omitted due to space constraint but it can be found in [15].

**Example**: For better insight about the usefulness of the proposed method, we provide numerical examples by employing the parameters used in the LTE network. For this network, different possible bandwidths are utilized (see [12] for more details). Here we consider the case where $B = 2.5$MHz with $T_d = 4.69\mu s$, coherence time $T_c = 0.5$ms, $T_p = \frac{T_c}{2}$ and $T_s = \frac{66.7}{256} = 0.2605\mu s$ [3], [12]. With these settings, we get

$$\tilde{K} = \frac{T_p}{T_d} \approx 53, \quad N_c = L = \frac{T_d}{T_s} \approx 18. \quad (16)$$

One can notice from (16) that our approach can serve $L\tilde{K} \approx 954$ UEs while mitigating pilot contamination. This is similar to scaling the number of UEs using the same resource by $L$. Note that if some resources are used for computational purpose, $\tilde{K}$ will be reduced slightly [3].

## VII. SIMULATION RESULTS

This section presents simulation results by considering a multicell system with $N_c$ cells where each cell serves $\tilde{K} = 4$ UEs. In this regard, we set $L = 4, g_{k1i} = 1, g_{k2i} = 0.9, g_{k3i} = 0.6, g_{k4i} = 0.7, \forall k, N_p = 16, M = 64, N = 2^{N_0}$ and $k = i = 1$ (i.e., the desired user and cell) with $N_0$ as an integer. We examine the average achievable rate of the UE's sub-carrier (i.e., $R_{kis}$). The SNR is defined as $\gamma = \frac{\mathrm{E}|d_{ki}|^2}{\tilde{\sigma}^2}, \forall k, i, \tilde{\sigma}^2 = \sigma^2$. All results are plotted by averaging 10000 channel realizations.

### A. Comparison of the Proposed and Existing Approaches

This subsection compares the performance of the proposed approach with those of the existing ones. In the proposed approach, first we choose $\mathbf{x}_{11}$ from a randomly selected unit energy quadrature phase shift keying (QPSK) signal. Then, we compute the remaining pilots just by shifting $\mathbf{x}_{11}$ appropriately with equally spaced conventional (fractional) discrete Fourier transform (DFT) vectors. For the existing approaches, we design $\mathbf{x}_{ki}, \forall i, k$ in two methods; the first method uses orthogonal pilots like in the proposed approach and the second method applies full pilot reuse (i.e., $\mathbf{x}_{ki} = \mathbf{x}_{kj}, \forall i, j, k$).

For fair comparison, we have utilized $\nu = 1$ (to resolve the multiplicative factor ambiguity) and binary phase shift keying transmitted signal of size 50 for the EDV approach of [7], and a full pilot reuse for [9]. Fig. 1 shows comparison of the proposed and existing approaches when $\gamma = 0$dB, $s = 20$ and $N_c = L$. From this figure, we can observe that utilizing orthogonal pilots slightly improve the performances of the MMSE and LS methods. Furthermore, the approaches of [7] and [9] also behave similar to the MMSE and LS approaches. However, the rate achieved by the proposed approach increases progressively with $N$. And, our approach achieves significantly higher $R_{kis}$ compared to those of the existing approaches especially in a massive MIMO regime which is desirable.

### B. Validation of Theoretical Results

This subsection validates the theoretical results provided in Sections V and VI. According to *Theorem 1*, if $N_c > 4$,

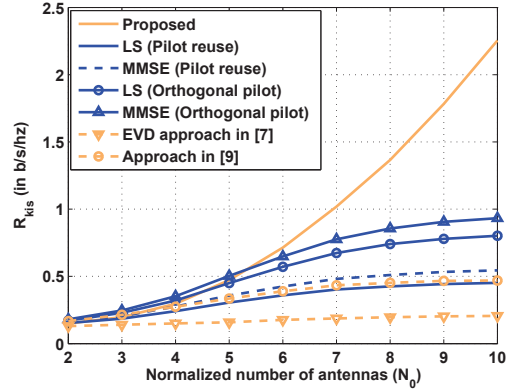

Fig. 1. Comparison of the rates achieved by the proposed and existing approaches.

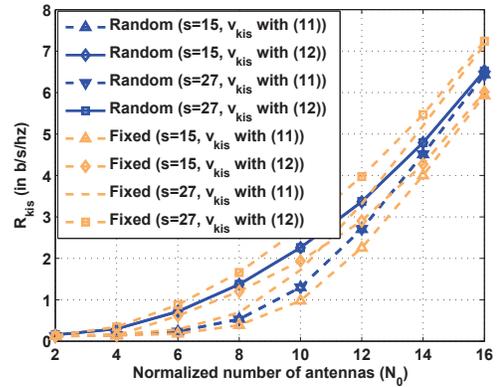

Fig. 2. Comparison of the rates achieved by fixed and random pilots.

increasing $N$ indefinitely does not help to improve $\bar{\gamma}_{kis}$. Also, the rate achieved by utilizing the solutions of (11) and (12) will be close to each other when $N$ is very large and $c_0$ of (15) is derived for $N \to \infty$. These motivate us to perform simulation for this subsection by considering very large $N$.

*1) Effect of Pilots:* This simulation examines the effect of pilots on the performance of the proposed algorithm. The pilot samples are designed by two approaches: The first approach utilizes $\mathbf{x}_{11} = N_p^{-1/2}(\mathbf{z}_1 + \sqrt{-1}\mathbf{z}_2)$ where $[\mathbf{z}_1 \ \mathbf{z}_2]$ are the last $2N_p$ samples generated from the "Maximum Length Sequence" $\pm 1$ bits of size 64 (i.e., fixed pilot assignment approach). The second approach designs the pilots like in Section VII-A (i.e., randomly selected pilot assignment approach). We examine the rate $R_{kis}$ (i.e., rate per user per sub-carrier) when $\gamma = 0$dB and $s = \{15, 27\}$ as shown in Fig. 2. As can be seen from this figure, in both sub-carriers, $R_{kis}$ increases with $N$ by utilizing either fixed or randomly generated pilots.

On the other hand, the average $R_{kis}$ may not be necessarily the same for all sub-carriers when fixed pilot assignment approach is utilized. However, the rate of these sub-carriers are the same when pilots are selected randomly which is expected (i.e., for given design approach of $\mathbf{v}_{kis}$). This simulation also exploits the fact that the rates achieved by employing the solutions of (11) and (12) are closer to each other when $N$ is very large. However, for small to medium $N$, higher $R_{kis}$ is obtained when $\mathbf{v}_{kis}$ is optimized using (12) which fits with the theory. In the following simulation, we utilize the $\mathbf{v}_{kis}$ designed by (12) with random pilot assignment approach.

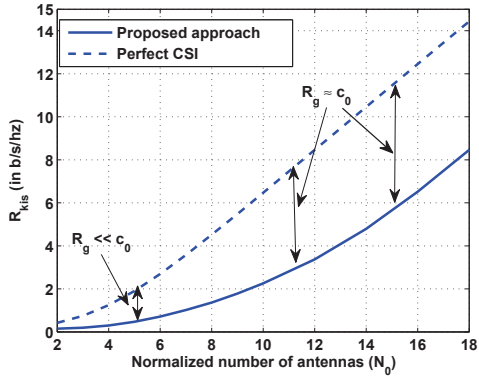

Fig. 3. The rates achieved by the proposed and perfect CSI scenarios.

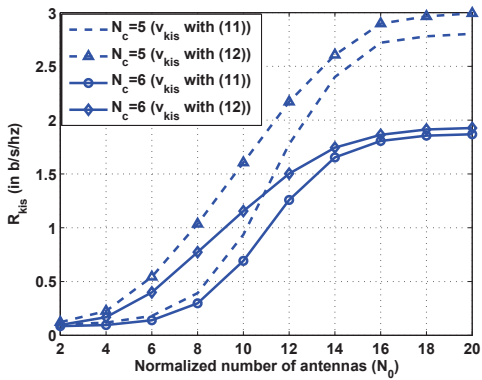

Fig. 4. Effect of $N_c$ for the proposed approach.

*2) Verification of $c_0$:* This simulation verifies the bound derived in (15) while utilizing $\gamma = 0$dB and $s = 20$ as shown in Fig. 3. From this figure, one can observe that the gap between the rate achieved by the proposed and that of perfect CSI scenario $R_g$ becomes constant ($R_g = c_0 \approx$ 6b/s/hz) after $N \approx 2^{16}$ which fits with the theory. On the other hand, $R_g$ decreases as $N$ decreases which is desirable in practice. For instance, if we deploy $N = 16$ antennas, the rate loss is $R_g <$ 1b/s/hz. We would like to mention here that, in a massive MIMO regime, we have observed almost the same behavior as in this figure for $\gamma = 20$dB. This is simply because the effect of noise vanishes for sufficiently large $N$ [3], [9].

*3) Effect of $N_c$:* As discussed in Section V, the proposed design experiences pilot contamination if $N_c > L = 4$. This simulation validates this claim. To this end, we set $s = 32$, $N_c = \{5, 6\}$, the channel gains of the UEs corresponding to the first four cells are the same as the first paragraph of this section, and the UEs of the fifth and sixth cells have channel gains $g_{k51} = 0.6, g_{k61} = 0.75, \forall k$. Fig. 4 shows the $R_{kis}$ achieved for these settings. As can be seen from this figure and Fig. 3, $R_{kis}$ drops quickly as we increase $N_c$ from 4 to $\{5, 6\}$. Furthermore, in both $N_c = \{5, 6\}$ cases, $R_{kis}$ is not increasing further after approximately $N = 2^{16}$ (i.e., bounded rate) which is inline with the theory provided in *Theorem 1*.

## VIII. CONCLUSIONS

This paper proposes novel joint channel estimation and beamforming approach for multicell wideband massive MIMO systems. Using the proposed approach, we determine $N_c$ utilizing the same time-frequency resource while mitigating the effect of pilot contamination for arbitrary channel covariance matrix of each UE. Specifically, when the channel has $L$ multipath taps, it is shown that $N_c \leq L$ cells can reliably estimate the channels of their UEs and perform beamforming while mitigating the effect of pilot contamination. Furthermore, whenever $N_c > L$, the rate of each UE will be bounded even if $N \to \infty$. These facts have been demonstrated using extensive computer simulations. The proposed joint channel estimation and beamforming approach is linear, simple to implement and significantly outperforms the existing approaches.